\documentclass[conference]{IEEEtran}
\IEEEoverridecommandlockouts
\usepackage{cite}
\usepackage{amsmath,amssymb,amsfonts}
\usepackage{algorithmic}
\usepackage{graphicx}
\usepackage{tabularx}
\usepackage{textcomp}
\usepackage{xcolor}
\usepackage{booktabs}
\usepackage{url}

\def\BibTeX{{\rm B\kern-.05em{\sc i\kern-.025em b}\kern-.08em
    T\kern-.1667em\lower.7ex\hbox{E}\kern-.125emX}}
\begin{document}

\title{A Reinforcement Learning Framework for Mobility Control of gNBs in Dynamic Radio Access Networks}

\author{
	\IEEEauthorblockN{
	Pedro Duarte, André Coelho, Manuel Ricardo}
	\IEEEauthorblockA{INESC TEC, Faculdade de Engenharia, Universidade do Porto, Portugal\\
    pedro.r.duarte@inesctec.pt, andre.f.coelho@inesctec.pt, manuel.ricardo@inesctec.pt}
 
    \thanks{This work was supported by the CONVERGE project which has received funding under the European Union’s Horizon Europe research and innovation programme under Grant Agreement No 101094831.}  
}

\maketitle

\begin{abstract}

The increasing complexity of wireless environments, characterized by user mobility and dynamic obstructions, poses challenges for the maintenance of Line-of-Sight (LoS) connectivity. Mobile base stations (gNBs) stand as a promising solution by physically relocating to restore or sustain LoS,  thereby necessitating the development of intelligent algorithms for autonomous movement control.

As part of the CONVERGE research project, which is developing an experimental chamber to integrate computer vision (CV) into mobile networks and enhance Quality of Service (QoS) in dynamic wireless environments, this paper presents two key contributions. First, we introduce the CONVERGE Chamber Simulator (CC-SIM), a 3D simulation environment for developing, training, and validating mobility control algorithms for mobile gNBs. CC-SIM models user and obstacle mobility, visual occlusion, and Radio Frequency (RF) propagation behavior.  It supports both offline reinforcement learning and real-time testing through tight integration with a standalone 5G system via the OpenAirInterface (OAI) RF simulator, enabling validation under realistic network conditions.

Second, leveraging CC-SIM, we develop a Deep Q-Network (DQN) agent that learns to reposition the gNB proactively in response to dynamic environmental changes. Experiments across three representative use cases show that the trained agent significantly reduces LoS blockage time—by up to 42\%—when compared to static deployments. These results highlight the effectiveness of learning-based mobility control in adaptive next-generation wireless networks.

\end{abstract}


\section{Introduction}\label{chap:1}

The rapid evolution of mobile networks, driven by increasing user demands, the proliferation of connected devices, and the rise of data-intensive applications, has pushed wireless infrastructure toward more advanced and adaptable designs. At higher frequencies---such as mmWave and sub-THz---performance relies on unobstructed Line-of-Sight (LoS), making real-world environments with user mobility and dynamic obstructions particularly challenging for maintaining link quality. One promising approach is the use of \textit{mobile base stations} (gNBs), which introduce physical adaptability to the radio access network (RAN) by repositioning in response to changing propagation conditions.

This work is conducted within the Horizon Europe CONVERGE project~\cite{converge_web}, which aims to develop tightly coupled radio and computer vision systems, enabling the network to ``view-to-communicate and communicate-to-view.'' A key objective is to enable mobile gNBs to dynamically reposition based on real-time perception of their surroundings within an experimental chamber.

This paper addresses the challenge of intelligently controlling the mobility of a gNB to mitigate LoS obstructions in dynamic environments. In dense urban or indoor scenarios, user equipment (UE) and obstacles may move unpredictably, requiring fast, context-aware decisions to preserve link quality. Traditional techniques such as handovers and beam switching are often reactive, introducing delays and performance degradation. In contrast, learning-based strategies that incorporate environmental awareness can proactively reposition the gNB to sustain communication performance.

The contributions of this work are twofold. First, we present the \textit{CONVERGE Chamber Simulator (CC-SIM)}, a 3D simulation environment that enables realistic modeling of spatial dynamics, visual occlusions, and wireless channel conditions. CC-SIM supports training, simulation, and live testing modes, and is tightly integrated with the OpenAirInterface (OAI) RF simulator to enable validation under realistic network conditions. Second, using CC-SIM, we design and train a Deep Q-Network (DQN) agent that learns to reposition the gNB in response to environmental changes, with the objective of maximizing LoS duration and minimizing path loss.

The remainder of this paper is organized as follows. Section~\ref{chap:cc_rl} presents the CONVERGE Chamber Simulator (CC-SIM) and the reinforcement learning approach used to train the gNB mobility controller. Section~\ref{chap:eval} describes the evaluation setup, experimental scenarios, and results. Section~\ref{chap:disc} discusses key insights and limitations. Finally, Section~\ref{chap:conc} concludes the paper and outlines directions for future work.
\section{Reinforcement Learning Framework for gNB Mobility Control} \label{chap:cc_rl}

Intelligent mobility control requires predictive, context-aware decision-making to proactively respond to evolving spatial conditions.
In order to enable such control, we formulate the problem as a reinforcement learning (RL) task, where an agent learns to move the gNB based on user and obstacle dynamics. This requires a large set of training samples capturing spatial interactions, visibility states, and their impact on radio link quality. While datasets like DeepSense6G~\cite{deepsense} and ViWi~\cite{viwi} offer valuable multimodal data for wireless and sensing tasks, they do not provide synchronized ground truth for gNB, UE, and obstacle motion jointly annotated with LoS and SNR metrics, which are essential for training and evaluating mobility control policies in dynamic environments.

To fill this gap, we developed the \textit{CONVERGE Chamber Simulator (CC-SIM)}: a lightweight, configurable simulation environment designed specifically for training and evaluating learning-based gNB mobility policies.

\subsection{CC-SIM}

CC-SIM emulates the spatial and RF behavior of a confined 3D environment where mobile network entities, such as a gNB, user equipment (UE), and dynamic obstacles, interact. The chamber is modeled as a bounded rectangular area with configurable dimensions and object definitions. Objects can follow either deterministic or randomized motion models, enabling the simulation of various LoS blockage scenarios.

Each entity's position is updated at discrete time steps, and the LoS visibility between the gNB and UE is computed using a geometric ray-tracing method that accounts for occlusions caused by obstacles. From this, a binary LoS indicator and a corresponding path loss value are derived. The path loss computation employs a simplified model that combines distance-based attenuation with a configurable attenuation factor for non-LoS (NLoS) conditions due to obstacles. This value can be optionally exported to the OAI RF simulator, supporting seamless integration with an emulated 5G network stack.

\begin{figure}[ht]
    \centering
    \includegraphics[width=\linewidth]{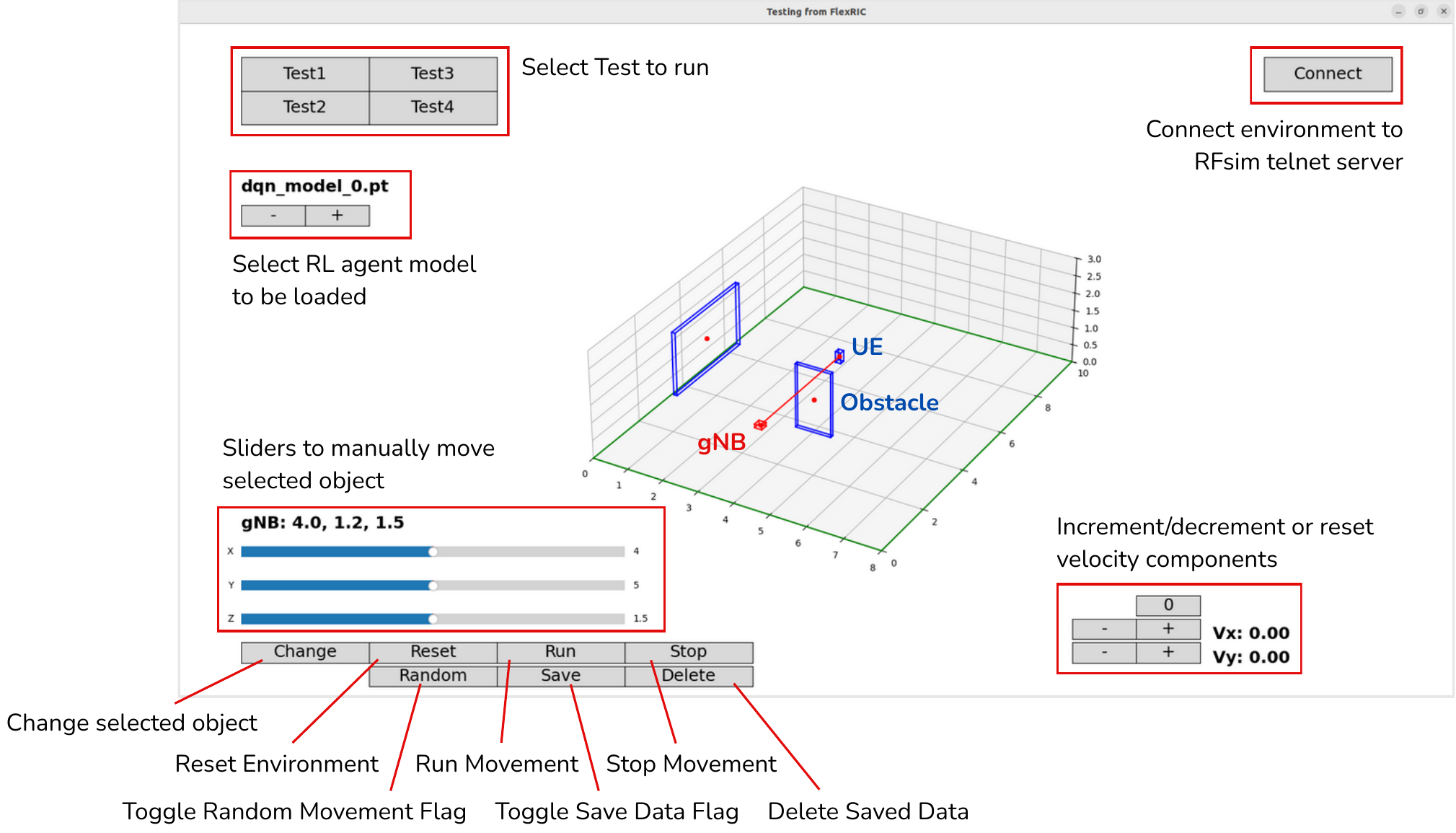}
    \caption{CC-SIM Graphical User Interface (GUI). The GUI enables interactive control over simulation parameters, including object velocity, trajectory patterns, and action execution. It offers real-time visibility feedback and supports telnet-based communication with the OAI RF simulator.}
    \label{fig:cc_gui}
\end{figure}

A graphical interface (GUI), shown in Fig.~\ref{fig:cc_gui}, facilitates visual debugging, manual control of object trajectories, and runtime adjustment of simulation parameters. The interface displays object positions, velocity vectors, and LoS status over time, aiding in the interpretation of the RL agent's behavior during training and evaluation.

CC-SIM operates in three modes: \textit{training}, used to collect environment feedback and reward signals for RL agent development; \textit{simulation}, used to evaluate learned policies offline; and \textit{live testing}, used to validate inference-based control when connected to a complete OAI-based Radio Access Network (RAN). In the following section, we describe the design of the RL agent and its interaction with the CC-SIM environment.

\subsection{RL Agent for gNB Mobility Control}

In order to enable intelligent gNB repositioning, we employ RL using a deep Q-network (DQN) to learn mobility policies that maintain LoS connectivity with the UE. The DQN algorithm \cite{dqn_algo} is well-suited for this task, as it supports sequential decision-making in dynamic environments and operates over discrete action spaces, thereby simplifying control and improving training stability compared to continuous action methods. 

\subsubsection{State Vector and Action Space}

In order to reduce the dimensionality of the problem and simplify training, the simulation scenario is restricted to a single UE and a single obstacle, with the gNB allowed to move only along the $x$-axis. Moreover, it is assumed that the gNB has perfect knowledge of the positions and velocities of all entities at each time step, enabling full observability for state construction. In practice, such information can be obtained through vision-based techniques combining object detection and motion estimation.

The agent's state at each time step consists of positional and kinematic features for the gNB, UE, and obstacle, along with a binary indicator for LoS obstruction, as summarized in Table~\ref{tab:inputvec}. These features form an 11-dimensional input vector, which is processed by a deep Q-network comprising three fully connected hidden layers with 64, 128, and 64 ReLU-activated neurons, followed by an output layer with three neurons corresponding to the Q-values of the discrete action set. The agent selects from this discrete action space, defined in Table~\ref{tab:actionspc}, by maintaining, incrementing, or decrementing its current velocity along the $x$-axis. Controlling velocity rather than absolute position facilitates smoother trajectory updates and reduces the likelihood of abrupt or erratic movements, thereby enhancing the convergence and stability of the mobility controller.
\vspace{-10px}
\begin{table}[ht]
\centering
\caption{State Vector Features.} \label{tab:inputvec}
\begin{tabular}{|p{2.5cm}|p{4.5cm}|}
\hline
\textbf{Feature} & \textbf{Description} \\
\hline
$x_{\text{gnb}}$ & gNB absolute \(x\)-position (m) \\
$x_{\text{gnb-ue}},~y_{\text{gnb-ue}}$ & Relative position: gNB to UE (m) \\
$x_{\text{gnb-obs}},~y_{\text{gnb-obs}}$ & Relative position: gNB to obstacle (m) \\
$v_{x_{\text{gnb}}}$ & gNB velocity in \(x\)-axis (m/s) \\
$v_{x_{\text{ue}}},~v_{y_{\text{ue}}}$ & UE velocity in \(x, y\) (m/s) \\
$v_{x_{\text{obs}}},~v_{y_{\text{obs}}}$ & Obstacle velocity in \(x, y\) (m/s) \\
$L_{\text{status}}$ & LoS status (0 = LoS, 1 = NLoS) \\
\hline
\end{tabular}
\vspace{10px}
\caption{Action Space.} \label{tab:actionspc}
\begin{tabular}{|p{2.5cm}|p{4.5cm}|}
\hline
\textbf{Action} & \textbf{Description} \\
\hline
0 & Maintain current velocity\\
1 & Increment velocity by $\delta$\\
2 & Decrement velocity by $\delta$\\
\hline
\end{tabular}

\end{table}

\subsubsection{Reward Function}

The reward guides the agent to preserve LoS and minimize path loss. When LoS is obstructed, a penalty in $[-1, 0[$ is applied, scaled by the normalized distance $\tilde{d}_{\text{oc}}$ between the LoS–obstacle intersection point and the obstacle’s center, encouraging early avoidance. If LoS is clear, the agent is rewarded based on proximity to the UE, using a squared normalized inverse mapping of the gNB–UE distance $d_{\text{ue}}$, scaled by a factor $k$ to adjust the reward magnitude, as summarized in Equation~\eqref{eq:reward}.

\begin{equation}\label{eq:reward}
R(s_t, a_t) = 
\begin{cases}
-1 + \tilde{d}_{\text{oc}}, & \text{if LoS obstructed} \\
k\cdot[\text{map}(d_{\text{ue}})]^2, & \text{otherwise}
\end{cases}
\end{equation}

\begin{figure}[ht]
    \centering
    \includegraphics[width=\linewidth]{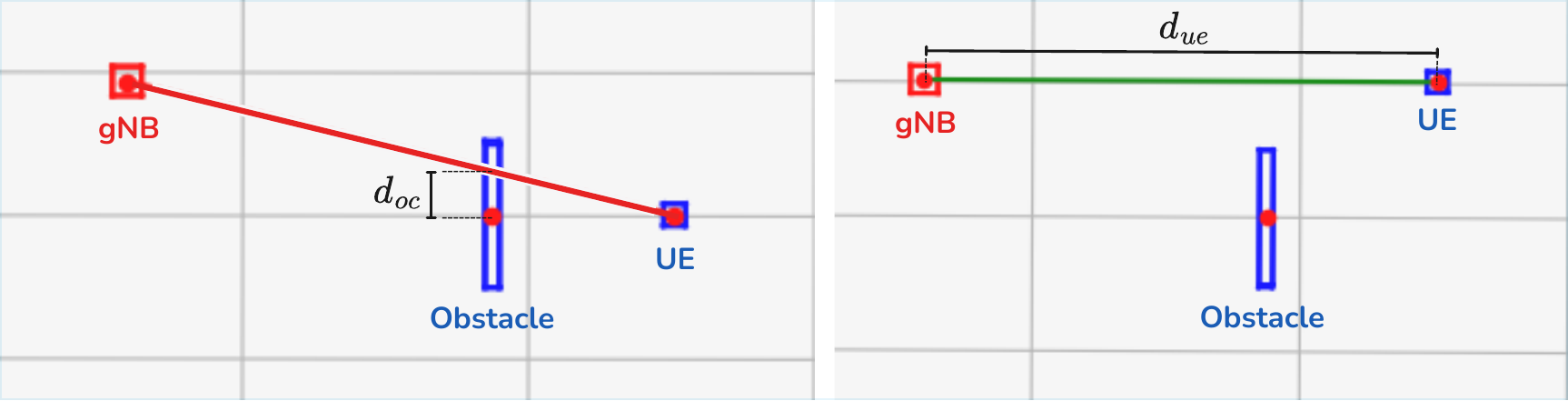}
    \caption{Reward function variables, with LoS obstructed (left) and LoS clear (right).}
    \label{fig:rl_var}
\end{figure}

Here, $\text{map}(\cdot)$ denotes a normalized inverse function bounded in $[0, 1]$. Both $\tilde{d}_{\text{oc}}$ and $d_{\text{ue}}$ are computed from $s_{t+1}$. Figure~\ref{fig:rl_var} illustrates the relevant function variables.

\subsubsection{Training Methodology}

The agent is trained using the DQN algorithm, based on the PyTorch tutorial by Zhou~\cite{zhouDQN}. Each training episode consists of four predefined mobility scenarios, each lasting a fixed number of time steps. This design exposes the agent to progressively complex and diverse environmental interactions within a single training session, while maintaining clear segmentation of behavioral contexts.

The four mobility scenarios are defined as follows: Scenario \textbf{A} -- a static obstacle positioned directly in front of the UE, representing a persistent LoS obstruction; Scenario \textbf{B} -- the obstacle moves laterally at a constant velocity, bouncing between chamber boundaries, while the UE remains stationary; Scenario \textbf{C} -- the obstacle is stationary and the UE moves in the same bouncing manner as described in Scenario B; Scenario \textbf{D} -- both the UE and obstacle move independently, bouncing asynchronously between boundaries at different speeds.

In all scenarios, the gNB is controlled by the RL agent.
The asynchronous movements of the UE and obstacle produce a variety of spatial configurations and LoS conditions, including simultaneous movements in the same or opposite directions. This diversity enhances the agent’s ability to generalize and adapt to varied UE-obstacle interaction patterns.
Scenario transitions occur after fixed step counts, allowing repeatable and interpretable progression through the training episode.

The agent is trained to perform control updates every 200\,ms, aligning with near-RT constraints, and action selection follows an \(\varepsilon\)-greedy policy, where \(\varepsilon\) decays linearly from 0.9 to 0.1.
Table~\ref{tab:episode_config} summarizes the main training parameters and episode layout, while Table~\ref{tab:ddqn_hparams} lists the DQN hyperparameters.

\begin{table}[ht]
\centering
\caption{Training parameters and episode configuration}
\label{tab:episode_config}
\begin{tabular}{lc}
\toprule
\textbf{Parameter} & \textbf{Value} \\
\midrule
Time step (\(\Delta t\)) & 0.2 s \\
Velocity step (\(\delta\)) & 0.35 m/s \\
Max gNB velocity (\(v_{\text{gNB}}^{max}\)) & 1 m/s \\
UE and obstacle velocity (\(v_{\text{obj}}\)) & 0.6 m/s \\
Number of training episodes (\(N\)) & 3 \\
Episode step limit (\(S_{\max}\)) & 3000 \\
Scenario A duration & 100 steps \\
Scenario B duration & 450 steps \\
Scenario C duration & 450 steps \\
Scenario D duration & 2000 steps \\
\bottomrule
\end{tabular}
\vspace{10px}
\caption{Training hyperparameters.}
\label{tab:ddqn_hparams}
\begin{tabular}{lc}
\toprule
\textbf{Parameter} & \textbf{Value} \\
\midrule
Batch size & 64 \\
Learning rate (\(\alpha\)) & 0.001 \\
Initial exploration rate (\(\varepsilon_\text{i}\)) & 0.9 \\
Final exploration rate (\(\varepsilon_\text{f}\)) & 0.1 \\
Discount factor (\(\gamma\)) & 0.9 \\
Target network update frequency & 100 steps \\
Replay buffer capacity & 1000 transitions \\
\bottomrule
\end{tabular}
\vspace{-15px}
\end{table}
\section{Evaluation} \label{chap:eval}

The evaluation setup consisted of the CC-SIM integrated with the OAI 5G RAN (gNB, UE, and 5GCore), where the trained DQN controlled the gNB movement. Traffic was generated between the 5GC and the UE to emulate realistic data flow, while CC-SIM provided real-time control, mobility modeling, and RF emulation to validate the RL-based controller in dynamic scenarios.

We conducted experiments across three use cases (UCs) and two movement patterns (MPs), defined as follows: \textbf{UC S} -- \textbf{s}tatic obstacle and \textbf{s}tatic UE; \textbf{UC O} -- mobile \textbf{o}bstacle and static UE; \textbf{UC U} -- mobile \textbf{U}E and static obstacle; \textbf{MP 1} -- mobile node moves from $x=2$~m to $x=6$~m at $v=0.6$~m/s; \textbf{MP 2} -- mobile node moves from $x=6$~m to $x=2$~m at $v=-0.6$~m/s.

A reference vision node was assumed to continuously capture all scene objects continuously for state estimation. The gNB was dynamically repositioned in real time, driven by near-RT inference from the RL agent.

Table~\ref{tab:use_cas} summarizes observed gNB behaviors for each use case, demonstrating adaptive control aligned with LoS maintenance. A video demonstration is available at \cite{video_demo}.

\begin{table}[!h]
    \centering
    \caption{gNB control behavior per use case.}
    \label{tab:use_cas}
    \vspace{-5px}
    \begin{tabular}{c p{6cm}}\\
\toprule
    \textbf{Use Case S} & gNB starts in a NLoS position and actively repositions to restore and maintain LoS connectivity.\\
\midrule
    \textbf{Use Case O} & gNB moves with the obstacle to avoid blockage and reverses direction when LoS risk increases, minimizing NLoS duration. \\
\midrule
    \textbf{Use Case U} & gNB initially moves opposite to the UE to preserve LoS, then tracks UE movement to reduce blockage.\\
\bottomrule
\end{tabular}
\end{table}

\begin{figure}[!h]
    \centering
    \includegraphics[width=\linewidth]{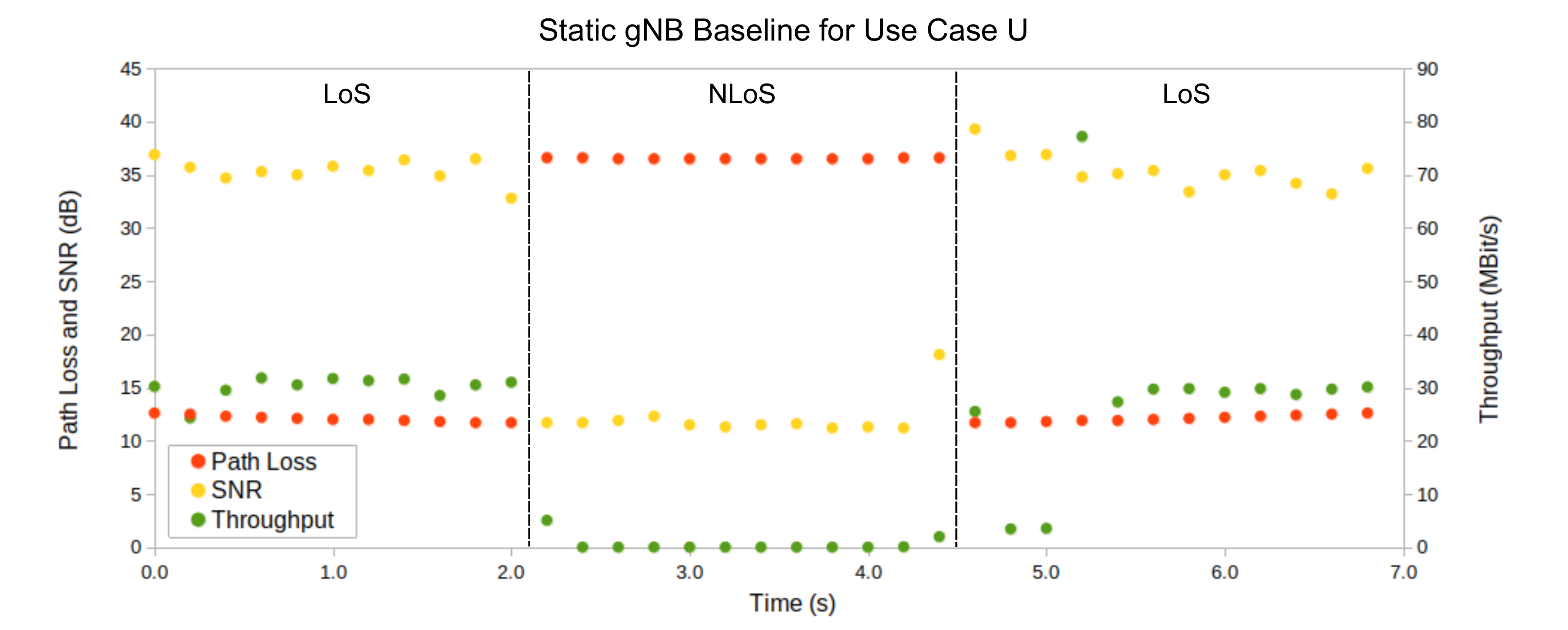}
    \includegraphics[width=0.9\linewidth]{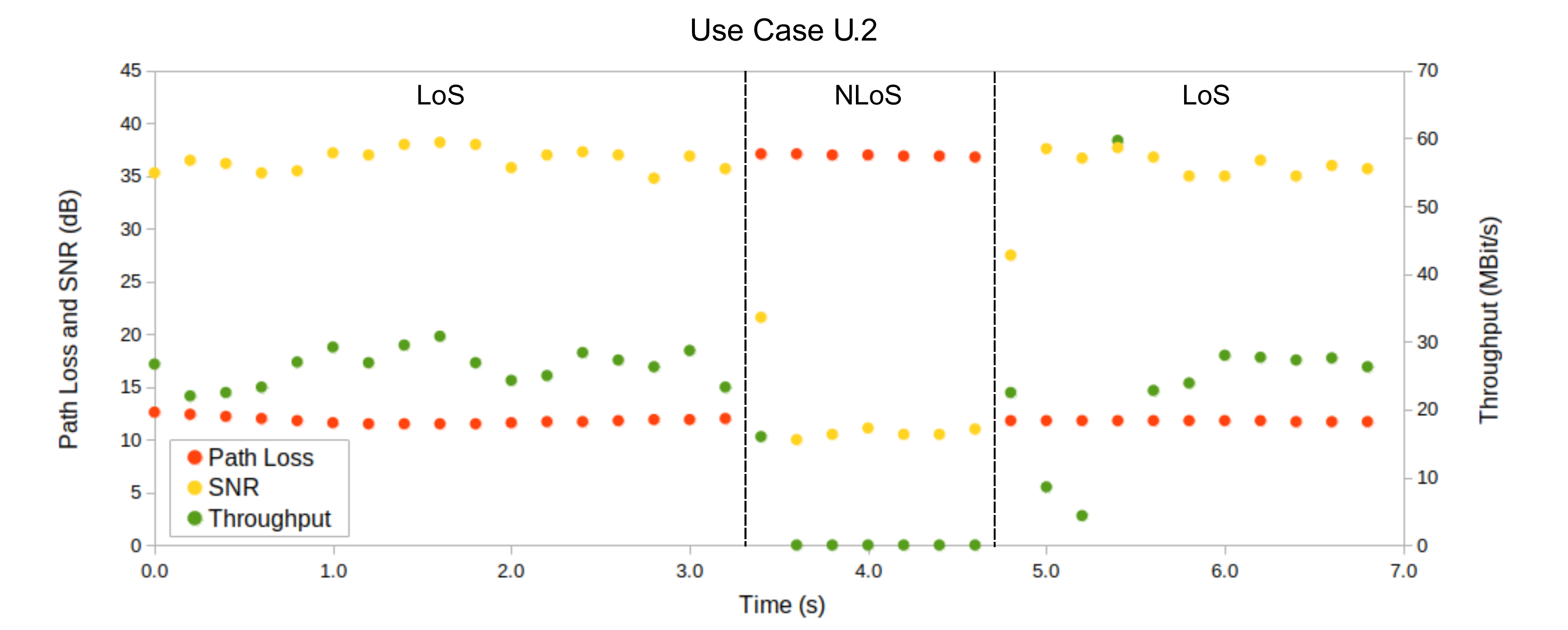}
    \caption[Use case U. Radio performance.]{Use Case U: Path loss, SNR, and throughput. Dynamic control reduces NLoS time by up to 41.6\%.}
    \label{fig:U_radio_results}
\end{figure}

Results from use cases O and U were compared to a static gNB baseline. Figure~\ref{fig:U_radio_results} compares path loss, SNR, and throughput, showing the static baseline in the top figure and the RL-controlled results in the bottom figure, demonstrating the best performance among all tests. The RL controller consistently delayed LoS obstruction onset, actively recovered connectivity, and significantly reduced total blockage time, as detailed in Table~\ref{tab:results}.

\begin{table}[!ht]
\centering
\caption{LoS obstruction delay and total NLoS time reduction relative to static gNB baseline}
\label{tab:results}
\begin{tabular}{lcc}
\toprule
\textbf{Test} & \textbf{Delay (s)} & \textbf{Time reduction (\%)} \\
\midrule
\textbf{O.1} & 1.0 & 12.5 \\
\textbf{O.2} & 0.8 & 25.0 \\
\textbf{U.1} & 1.6 & 25.0 \\
\textbf{U.2} & 1.2 & \textbf{41.6} \\
\bottomrule
\end{tabular}
\end{table}
\section{Discussion} \label{chap:disc}

The experimental evaluation confirmed that the RL-based controller maintained LoS and link quality by adaptively repositioning the gNB using spatial data. Although tested in a simplified scenario, the controller successfully demonstrated the feasibility of the proposed framework. Performance gains over the static baseline validated both the reliability of CC-SIM as a development and test environment, as well as the effectiveness of learning-based approaches for intelligent gNB mobility control.

Although deployed in simulation, the trained controller can be exported to real gNB control applications where environmental object positions can be obtained through CV and sensing data. Replacing the simulator with real sensor inputs enables the direct deployment of learned policies in physical testbeds. Furthermore, CC-SIM has potential as a digital twin for real-time mirroring and monitoring of deployed networks.

The controller was trained assuming full observability via a reference vision node. While this supported rapid prototyping, real-world scenarios will require policies capable of handling partial observability. Approaches such as recurrent neural networks or attention mechanisms can address uncertainty. In such cases, RF measurements can provide complementary input, enhancing the multimodal nature of control.

This work focused on a single UE and one-dimensional gNB movement, but the underlying architecture supports scalable, more complex scenarios, which may require extending the control logic for operation in richer environments.

The obtained results demonstrate the viability of intelligent mobility control for mobile gNBs and highlight the benefits of modular, data-driven frameworks for future wireless network management.
\section{Conclusions} \label{chap:conc}

This paper addresses the challenge of controlling the position of a mobile gNB in real time to maintain LoS connectivity with a UE in dynamic environments. We formulated the problem as an RL task and developed the CONVERGE Chamber Simulator (CC-SIM), a configurable 3D simulation environment for training and evaluating intelligent mobility control policies.

Using CC-SIM, we trained an RL-based controller that adaptively repositions the gNB based on spatial information about the user and obstacles. The obtained results demonstrate that the controller maintained LoS, and improved link quality and network performance, reducing blockage duration by up to 41.6\% compared to a static baseline. These outcomes present a promising research direction for the intelligent control of mobile base stations in future wireless networks.

Future work will focus on extending the controller to handle multi-object scenarios and more complex mobility, as well as developing an experimental dataset comprising diverse scenarios with object positions, velocities, and RF metrics. This will support broader training, benchmarking, and testing of alternative control gNB techniques and ML algorithms for practical deployment in dynamic wireless networks.

\end{document}